\documentclass[aps,prl,twocolumn,showpacs,superscriptaddress]{revtex4}
\usepackage{graphicx}

\begin{document}

\title{Quantum Friction of Micromechanical Resonators at Low Temperatures}

\author{Kang-Hun Ahn}
\affiliation{
School of Physics, Seoul National University, Seoul 151-747, Korea 
}
\author{Pritiraj Mohanty}
\affiliation{
 Department of Physics, Boston University,  Boston, MA 02215 
}
\date{ \today}

\begin{abstract}
Dissipation of micro- and nano-scale mechanical structures is dominated
by quantum-mechanical tunneling of two-level defects intrinsically present 
in the system. We find that at high frequencies---usually, for smaller, micron-scale 
structures---a novel mechanism of phonon pumping of two-level defects gives 
rise to weakly temperature-dependent internal friction, $Q^{-1}$, 
concomitant to the effects observed in 
recent experiments. Due to their size, comparable to or shorter than the emitted 
phonon wavelength, these structures suffer from superradiance-enhanced 
dissipation by the collective relaxation of a large number of two-level defects
contained within the wavelength.
\end{abstract}
\maketitle


Physical properties of physically- and chemically-engineered micromechanical systems
are of immense fundamental and technical interest \cite{physics-today}.
 Some of the recent spectacular examples of micromechanical structures include the measurement of
 vortex motion in high-{$T_c$} superconductors \cite{bishop-nature}, biomolecular recognition \cite{biomechanics}, 
actuation of sensors via Casimir force \cite{casimir}, and shuttling of electron charge in a quantum dot \cite{erbe}. 
Important to all these experiments is the oscillation of a particular or a set of micron-sized resonators 
at a resonance frequency determined by the geometry and material properties. 
Changes in the "resonant" oscillation frequency or the oscillation amplitude mostly
 determine the magnitude of the force of interest, to which the micromechanical structure is coupled. 
Detrimental to the detection of force is the damping of the resonant structure, quantified by 
quality factor Q or dissipation $Q^{-1}$. 

The essential problem inherent to the types of force measurement mentioned above is
the low quality factor $Q$, observed in small resonators. In many experiments, 
intrinsic two-level defects\cite{phillips2,esquinazi} are found to be the dominant
cause of internal friction $Q^{-1}$ in crystalline resonators \cite{kleiman,mihailovich,mohanty}.
The observed linear temperature dependence of $Q^{-1}\propto T$ \cite{kleiman} could be 
explained assuming reasonable density of defects within a linear response theory of 
two-level systems (TLS)\cite{phillips,keyes}. 
Despite the success of the TLS mechanism, dissipation at low temperatures continues 
to hold many challenges, notably the observation of a weak temperature dependence \cite{mohanty} 
and the non-monotonic dependence of $Q^{-1}(T)$ in silicon resonators \cite{greywall}. 
Most recent experiments merely add to the list of problems yet to be understood.

Previous theoretical works on the acoustic response of TLS use 
the linear response approximation or adiabatic approximation for nonlinear 
response\cite{galperin,stockburger}. The question we would like to explore is 
whether the anomaly can be explained in terms of TLS if both assumptions are
abandoned. For dissipation in micromechanical resonators, such issues become 
terribly important, as a new parameter space emerges with decreasing size.
For example, emitted phonon wavelength 
 becomes much longer ($\gg 1 \mu m$) compared to the system size at low temperatures, bringing
novel correlation effects\cite{tobias}, extremely relavant to mesoscopic structures.

In this work, we evaluate internal friction $Q^{-1}$ when the 
two-level defects operate non-linearly and non-adiabatically. We show that the 
nonlinear response of $Q^{-1}$ results in a large contribution at low temperatures 
at high driving frequencies due to the process of {\em phonon pumping}.  
We propose and demonstrate a mechanism of how $Q^{-1}$ can be significantly enhanced 
by the phonon pumping process through {\em cooperative emission} of phonons. Cooperative
emission becomes possible because of the long wavelength of the emitted phonons, which allows
the correlated decay of two-level systems contained within the wavelength. Our theory
provides a mechanism for enhanced dissipation in micromechanical resonators within
the context of two-level systems.

The Hamiltonian of a TLS is fully characterized by the asymmetry energy $\Delta$  and
the tunneling matrix element $\Delta_{0}$. 
The  asymmetry energy of a TLS is the energy splitting due to 
the static strain  $\pm \frac{1}{2}\Delta_{s}$ and
the applied time-dependent strain field $\epsilon_{0}\cos\omega t$: 
$ \Delta(t)=\Delta_{s}+2\gamma\epsilon_{0}\cos\omega t$, where 
$\gamma$ is the constant of coupling between the acoustic wave and the asymmetry energy.
Friction due to thermal phonons is described by the Hamiltonian for the 
single TLS coupled to phonon fields\cite{esquinazi};
\begin{eqnarray}
\nonumber
H&=&\frac{\Delta_{s}+2\gamma\epsilon_{0}\cos\omega t}{2}\sigma_{z}
+\frac{\Delta_{0}}{2}\sigma_{x}+\sum_{{\bf q},\alpha}
\hbar \omega_{\alpha}({\bf q})
a^{\dagger}_{{\bf q},\alpha} a_{{\bf q},\alpha} \\
&+&\sum_{{\bf q},\alpha}\gamma_{\alpha}
\sqrt{\frac{q\hbar}{2\rho {\mathcal V}v_{\alpha}}}\sigma_{z}
(a_{{\bf q},\alpha}+ a^{\dagger}_{-{\bf q},\alpha}),
\label{h1}
\end{eqnarray}
where 
 ${\mathcal V}$ and $\rho$ are the volume and the mass density of the resonator,
$v_{\alpha}$ is the sound velocity for polarization $\alpha$,
$a_{{\bf q},\alpha}
(a^{\dagger}_{{\bf q},\alpha})$ is the phonon annihilation(creation) operator.
 $\gamma_{\alpha}$ is the coupling constant which relates the change in the asymmetry
energy to the local elastic strain.

In crystalline materials with many TLSs, it is known that the static asymmetry 
energy $\Delta_{s}$ has a wide distribution due to the randomness of local strain, 
while tunneling splitting energy $\Delta_{0}$ has a well-defined value\cite{phillips}.
For simplicity, we consider a 
rectangular distribution\cite{keyes} of $\Delta_{s}$ with $n$ number of TLS per unit volume 
in $0<\Delta_{s} < \Delta_{s}^{max}$. The underlying assumption,
$\hbar \omega \ll \gamma \epsilon_{0} \ll \Delta_{s}^{max}$, holds in most experiments.

The internal friction $Q^{-1}$ is related to the  
energy loss per unit volume $\delta E$ in a cycle of the acoustic wave ;
$Q^{-1}=\frac{\delta E }{2\pi E_{0}}$,
where $E_{0}=2\rho \epsilon_{0}^{2}v^{2}$ is the acoustic wave energy stored per unit volume\cite{esquinazi}.
The time-averaged phonon emission power $\bar{P}(\Delta_{s},T)$ of a TLS is 
related to  $Q^{-1}$: 
$Q^{-1}(T)=\frac{n}{2\rho v^{2}}\frac{1}{\epsilon^{2}_{0}\omega}\int_{0}^{\Delta_{s}^{max}} 
\bar{P}(\Delta_{s},T) d\Delta_{s}.$
Furthermore, we divide the two-level systems into two kinds based on their asymmetry energy
$\Delta_{s}$: (i) $\Delta_{s} < 2\gamma\epsilon_{0}$, and (ii) $\Delta_{s} > 2\gamma\epsilon_{0} $; which contribute
to the total dissipation according to $Q^{-1}(T)=Q^{-1}_{1}+Q^{-1}_{2}$:
\begin{eqnarray}
Q^{-1}_{1}(T)&=&\frac{n}{2\rho v^{2}}\frac{1}{\epsilon^{2}_{0}\omega}\int_{0}^{2\gamma\epsilon_{0}} 
\bar{P}(\Delta_{s},T) d\Delta_{s}
\label{q1}
\\ 
Q^{-1}_{2}(T)&=&\frac{n}{2\rho v^{2}}\frac{1}{\epsilon^{2}_{0}\omega}\int_{2\gamma\epsilon_{0}}^{\Delta_{s}^{max}}
 \bar{P}(\Delta_{s},T) d\Delta_{s}. 
\label{q2}
\end{eqnarray}

\noindent
In linear response theory where $\epsilon_{0}$ is infinitesimally small, all the TLS 
belong to category (ii), and $Q^{-1}_{1}=0$. However, even a small experimental 
value of $\epsilon_{0}=10^{-6}$ will give rise to nonlinear behavior\cite{kleiman}, which 
implies that TLS with $\Delta_{s} < 2\gamma\epsilon_{0}$ may also play an important role\cite{keyes}.

Usually a time-dependent unitary transformation $U$ is used for solving time-dependent 
Hamiltonian in the nonlinear regime, so that the TLS has time-dependent energy levels 
given as $\pm\frac{1}{2}E(t)$,($ E(t)=\sqrt{\Delta(t)^{2}+\Delta_{0}^{2}}$). 
In the standard adiabatic limit\cite{galperin,stockburger} used to explain 
the low-frequency experiments, the acoustic field frequency is assumed to be 
very low so that the additional term due to the unitary transformation
$-i\hbar U^{\dagger}\frac{\partial U}{\partial t }= \sigma_{y}
\frac{\Delta_{0}\gamma\epsilon_{0}}{E(t)^{2}}\hbar \omega \sin \omega t  $ can be 
safely neglected.
In contrast, here we consider the high frequency regime where the matrix element of this additional 
term can not be neglected in comparison to the diagonal term $\frac{1}{2}E(t)\sigma_{z}$ even at 
the level-crossing point $E(t)=\Delta_{0}$:
\begin{eqnarray}
\frac{\gamma\epsilon_{0}\hbar\omega}{\Delta_{0}} \gg \Delta_{0}.
\label{highf}
\end{eqnarray}
Under this condition, as the tunneling motion between the two sites is not fast compared 
to the rapid time evolution of the acoustic wave, it is not useful to take the TLS 
 bonding states of the two sites as the basis states. Instead, it is more natural to consider 
the localized basis states in each local potential minimum and consider the tunneling 
term $\frac{1}{2}\Delta_{0}\sigma_{x}$ as a perturbation.

We treat the coupling to the acoustic wave adiabatically and treat the tunneling term and 
the coupling to the thermal phonon pertubatively. 
The approximation is guaranteed by the fact TLS-phonon interaction can be treated pertubatively 
whenever the two-level approximation is valid\cite{leggett}. Let $|l>$ and $|r>$ be the left and 
the right localized states respectively.
The leading term for the transition from one state to the other is the second order process 
involving the tunnelling and emission (absorption) of a phonon; at low temperatures single phonon process dominates. 
The thermal averaged relaxation rate is written as
\begin{eqnarray}
\frac{1}{\tau(\Delta)}=
 A\Delta_{0}^{2} \Delta\coth\big(\frac{\Delta}{2k_{B}T}\big),
\end{eqnarray}
where $A=\sum_{\alpha}\frac{\gamma^{2}_{\alpha}} {2\pi \hbar^{4}v^{5}_{\alpha}\rho} $.
Here we assume the Debye density of states of phonon  
with a linear dispersion: $\omega_{\alpha}=v_{\alpha}q $.

Time evolution of the occupation number $n_{r}(=1-n_{l})$ follows the rate equation\cite{galperin} 
for the time-dependent asymmetry energy $\Delta(t)=2\gamma\epsilon_{0}\cos\omega t +\Delta_{s}$:
\begin{eqnarray}
\frac{d n_{r}}{dt}=-\frac{n_{r}-n_{r}^{(0)}}{\tau(\Delta(t))}; ~~~~n_{r}^{(0)}(t)=1/(e^{\beta\Delta(t)}+1),
\label{rate1}
\end{eqnarray}
where $\beta=1/k_{B}T$.
The typical process is depicted in Fig.~1.
In the regime $ \Delta_{s} < 2\gamma\epsilon_{0} $, the two energy levels cross, and at this point population
inversion arises. 
Note that the population inversion process ( $A\rightarrow B \rightarrow C$ )
is followed by the spontaneous emission of phonon ($C\rightarrow D$). 
The high-frequency assumption of Eq.(\ref{highf}) is crucial for the pumping process near the
level-crossing point. In the conventional adiabatic limit 
($\gamma\epsilon_{0}\hbar\omega/\Delta_{0} \ll \Delta_{0}$) with TLS energies of $\pm\frac{1}{2}E(t)$, 
level crossing is avoided and the TLS in the low (high) energy state remains mostly 
in the low(high) energy state.

\begin{figure}
\includegraphics[width=8.0cm,height=7cm]{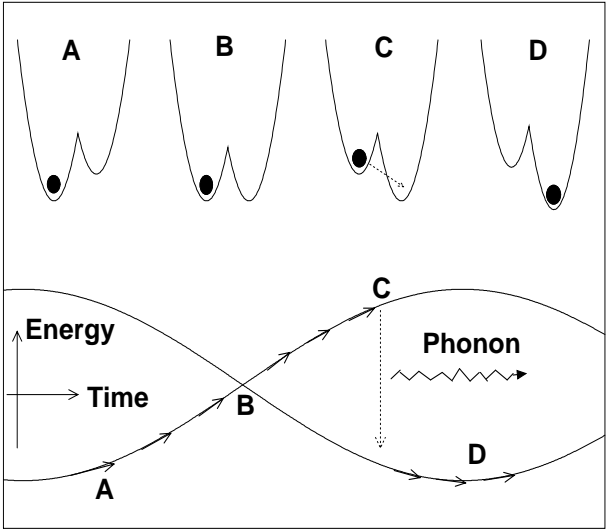}
 \caption{The schematic diagram describes the energy-pumping process of the
two-level atom from the acoustic wave to phonons. 
The crossing curves represent the time evolution of the two energy levels 
with the energy gap of $\Delta(t)=2\gamma\epsilon_{0}\cos\omega t+\Delta_{s}$. 
The energy diagrams of the two-level atom  as a function of the configuration coordinates are depicted in the upper panel. }
\label{pumping}
\end{figure}

The time-averaged phonon emission power\cite{galperin} can be obtained by solving (Eq.\ref{rate1}) for $n_{r}(t)$;
\begin{eqnarray}
\bar{P}(\Delta_{s},T)=-\frac{\omega}{2\pi}\int_{0}^{\frac{2\pi}{\omega}} dt \Delta(t) \frac{dn_{r}(t)}{dt}.
\label{barp}
\end{eqnarray}
For the general case of asymmetric TLS, numerical integration of Eq.(\ref{rate1}) 
gives $\bar{P}$ as a function of the static asymmetry energy $\Delta_{s}$ and temperature $ T$, 
and Eq.(\ref{q1}) subsequently gives $Q^{-1}_{1}$. For the particular case of symmetric 
TLS ($\Delta_{s}=0$), Eq.(\ref{barp}) can be rewritten using an analytic solution of 
$n_{r}(t)$ at $T=0$;
\begin{eqnarray}
\bar{P}=
\frac{2\gamma^{2}\epsilon_{0}^{2}A\Delta_{0}^{2}}{\pi\omega
}
\int^{\frac{\pi}{2\omega}}_{-\frac{\pi}{2\omega}}\cos^{2}\omega t
\frac{\exp(\frac{\sin\omega t}{\omega \tau_{\epsilon}})}
{\cosh(1/\omega\tau_{\epsilon})}
 dt
\label{pv0}
\end{eqnarray} 
where $\tau_{\epsilon}$ is a typical 
relaxation time scale of symmetric TLS at zero temperature 
when $\omega\tau_{\epsilon} >1$;
\begin{eqnarray}
\frac{1}{\tau_{\epsilon}}=2A\Delta_{0}^{2}\gamma\epsilon_{0}.
\label{taue}
\end{eqnarray}
When $\Delta_{s} > 2\gamma\epsilon_{0}$, the TLS always stays in the lower energy state at zero temperature, 
hence the TLS does not emit a phonon in its steady state, $\bar{P}(\Delta_{s} >2\gamma\epsilon_{0}, T=0 )= 0$. 
Since $\bar{P}(\Delta_{s},T=0)$ is a smoothly-decaying function of $\Delta_{s}$,
 an analytic formula for $Q_{1}^{-1}(T=0)$
can be obtained using the approximation 
$\int_{0}^{ 2\gamma\epsilon_{0}}\bar{P}(\Delta_{s},T=0)d\Delta_{s}\approx 
\gamma\epsilon_{0}(\bar{P}(0,0)+\bar{P}(\gamma\epsilon_{0},0))
= \gamma\epsilon_{0}\bar{P}(0,0)$.
From Eq.(\ref{pv0}), the zero temperature internal friction $Q^{-1}_{1}$ is found to be
\begin{eqnarray}
Q_{1}^{-1}(T=0)\approx\frac{n\gamma^{2}}{\rho v^{2}}~ {\rm sech}(\frac{1}{\omega\tau_{\epsilon}})
I_{1}(\frac{1}{\omega\tau_{\epsilon}}),
\label{qq1}
\end{eqnarray}
where $I_{1}(x)$ is the first kind modified Bessel function of order 1. 

Now let us turn to $Q_{2}^{-1}$.
For weak strain $2\gamma\epsilon_{0} < \Delta_{s} $, where the energy levels do not cross,
phonon pumping process does not take place. 
This is the regime where the usual linear response theory is a good approximation for describing internal friction. 
In the linear response theory\cite{phillips,phillips2}, $Q_{2}^{-1}(T)$ in Eq.(\ref{q2}) 
for $k_{B}T \gg \hbar\omega >  \hbar/\tau(\Delta_{0}) $  can be written as\cite{phillips2}
\begin{eqnarray}
Q_{2}^{-1}(T)=\frac{n\gamma^{2}}{\rho v^{2}k_{B}T}\frac{1}{\omega}
\int_{2\gamma\epsilon_{0}}^{\Delta_{s}^{max}}\frac{\Delta^{2}}{E^{2}}{\rm sech}^{2}(\frac{E}{2k_{B}T})\tau^{-1}(E)d\Delta,
\end{eqnarray}
where $E=\sqrt{\Delta^{2}+\Delta_{0}^{2}}$.
For $\Delta_{0} \ll \gamma\epsilon_{0} \ll \Delta_{s}^{max}$, an expression for $Q^{-1}_{2}$ 
is easily found:
\begin{eqnarray}
Q_{2}^{-1}(T) = \frac{n\gamma^{2}}{\rho v^{2}}\frac{ 2 }{\omega\tau_{\epsilon}}
f\big(\frac{k_{B}T}{2\gamma\epsilon_{0}}\big),
\label{q22}
\end{eqnarray}
where $f(x)$ is an almost linear function, defined by
$f(x)=x\int_{1/x}^{\infty}dt\frac{t}{\sinh t}$.

Fig. 2 shows the temperature dependence of $Q^{-1}=Q_{1}^{-1}+Q_{2}^{-1}$ for
$\omega\tau_{\epsilon}=1$, calculated from Eqs.(\ref{q1},\ref{q2},\ref{rate1},\ref{barp} and \ref{q22}). 
$Q^{-1}$ shows a weak temperature dependence with decreasing temperature, 
as $Q_{1}^{-1}$ becomes more important than $Q_{2}^{-1}$ at low temperatures.
The low-temperature saturation value of $Q^{-1}$ for $\omega\tau_{\epsilon} \ge 1$ is well approximated by 
 $\sim 0.3 (1/\omega\tau_{\epsilon})( n\gamma^{2}/\rho v^{2})$. 
Now, let us estimate $\tau_{\epsilon}$ from experimental data\cite{kleiman}.
Since $n\gamma^{2}/\rho v^{2}$ is roughly the high-temperature saturation\cite{phillips2} value of $Q^{-1}$,
for the single-crystalline silicon data in Ref.\cite{kleiman}, 
we estimate $n\gamma^{2}/\rho v^{2}\sim 10^{-6}-10^{-5}$ from the experiments at $\omega \sim 10^{3} -10^{4}$Hz.
Since the low temperature saturation value of $Q^{-1}$ was found to be $\sim 10^{-7} -10^{-6}$ for these frequencies,
we estimate $\omega\tau_{\epsilon}\sim 1$, which means $\tau_{\epsilon}\sim 10^{-3}-10^{-4}$sec.
However, these values  of  $\tau_{\epsilon}$ are  anomalously small compared 
to the direct estimate using Eq.(\ref{taue})with reasonable values of parameters\cite{phillips,phillips2,esquinazi}.
By using $\gamma\sim 1 eV,\epsilon_{0}\sim 10^{-6}, A=10^{8}s^{-1}k_{B}^{-3}$ into Eq.(\ref{taue}), 
we get $\tau_{\epsilon}\sim 1$ sec.
In other words,  the low-temperature dissipation in experiments\cite{kleiman} is anomalously 
larger than our direct theoretical estimation.
\begin{figure}
\includegraphics[width=8.0cm,height=7cm]{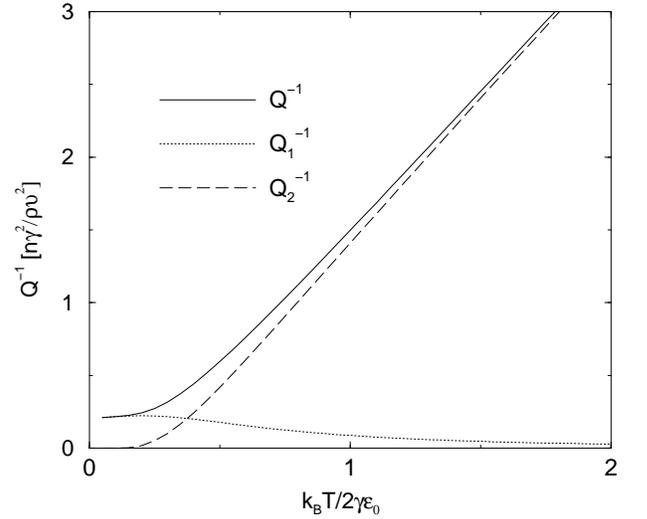}
 \caption{ The internal friction $Q^{-1}=Q_{1}^{-1}+Q_{2}^{-1}$ as a function of
temperature for $\omega\tau_{\epsilon}=1$.}
\end{figure}

Here we propose and discuss the origin of the anomalously small relaxation time $\tau_{\epsilon}$. 
We believe that the {\it cooperative emission} \cite{dicke} of phonons provides a natural 
mechanism for the short relaxation time. In Dicke superradiance \cite{dicke},
$N$ number of two-level atoms within a distance shorter than the emitted photon wavelength emit 
photons cooperatively, therefore the lifetime 
of the atom in the excited state is effectively reduced by a factor of $N$.
In the usual treatment of phonon emission of TLS, as we have done so far, 
the emission of phonon is calculated by assuming that the TLSs are totally independent.
However, this assumption is not valid when a number of TLS are contained within 
a distance shorter than the wavelength of the emitted phonon. This phonon wavelength,
$\lambda_{ph} \sim {\hbar v_s \over \gamma\epsilon_{0}}$ ($v_{s}$ is the sound velocity),
 can be considerably larger ($\stackrel{>}{\sim} \mu m$), leading to the cooperative emission from a large number of two-level systems.

Now let us consider $N$ two-level systems within a volume $\lambda_{ph}^{3}$. 
In the context of phonon emission, pumping of the two-level system is
done by the acoustic wave applied to the resonator.
The many-TLS Hamiltonian in Eq.(\ref{h1}) can be generalized to 
 $H=H_{0}+H_{tun}+H_{TLS-ph}$ where
$H_{0}=H_{TLS}+H_{ph}=
\frac{1}{2}\Delta(t)\sum_{i=1}^{N}\sigma_{i,z}
+\sum_{{\bf q},\alpha}
\hbar \omega_{\alpha}({\bf q})
a^{\dagger}_{{\bf q},\alpha} a_{{\bf q},\alpha}$, 
$H_{tun}= \frac{\Delta_{0}}{2}\sum_{i=1}^{N}\sigma_{i,x}$, and 
\begin{eqnarray}
H_{TLS-ph}=\sum_{i=1}^{N}\sum_{{\bf q},\alpha}\gamma_{\alpha}
\sqrt{\frac{q\hbar}{2\rho {\mathcal V}v_{\alpha}}}\sigma_{i,z}
(a_{{\bf q},\alpha}+ a^{\dagger}_{-{\bf q},\alpha}).
\end{eqnarray}
 
The many-TLS state $\big|l,l,l,...l\big>$, where all the TLS are in the $l$ state, is the maximally 
excited state or the ground state.
We consider the set of many-TLS states 
generated from $H_{TLS-ph}+H_{tun}$, which are
\begin{eqnarray}
\big|M\big>= \sqrt{\frac{(N/2-M)!}{N!(N/2+M)!}}
\left(\sum_{i=1}^{N}\sigma_{i}^{+}\right)^{N/2+M} \big|l,l,l,...l\big>,
\end{eqnarray}
where $M=-N/2,-N/2+1,-N/2+2,...,N/2$.
These states satisfy $H_{TLS}(t)\big| M \big>=M\Delta(t)\big| M\big>$.
As in the case of a single TLS, a perturbation analysis 
finds the transition rate $w_{M\pm1,M}$  from  $\big| M \big>$ to $\big| M \pm 1 \big>$:
to be 
\begin{eqnarray}
w_{M\pm1,M}(t)=w(\mp \Delta(t)) \left( \frac{N}{2} \mp M  \right) 
\left(\frac{N}{2}\pm M+1\right),
\end{eqnarray}
where 
\begin{eqnarray}
w(\Delta)= A\Delta_{0}^{2} \frac{\Delta}{1-\exp({-\Delta/k_{B}T})}
\end{eqnarray}
Here, we have used the relation
$\big< M \pm 1\big| \sum_{i=1}^{N}\sigma^{\pm}_{i} \big| M\big>
=\sqrt{\big(N/2\mp M)\big(N/2\pm M+1)}$. 
Now the following rate equation for the many-TLS remains to be solved numerically:
\begin{eqnarray}
\nonumber
\frac{d n_{M} }{dt}&=& 
 w_{M,M+1}(t)n_{M+1} -w_{M-1,M}(t)n_{M} 
\\
&+&w_{M,M-1}(t)n_{M-1} -w_{M+1,M}(t)n_{M},
\label{rate2}
\end{eqnarray}
The time-averaged power $\bar{P}_{N}$ emitted by $N$ number of TLS
can be obtained as: 
\begin{eqnarray}
\bar{P}_{N}=- \frac{\omega}{2\pi} \int_{0}^{2\pi/\omega} dt
  \sum_{M=-N/2}^{N/2} M \Delta(t)\frac{d n_{M}(t)}{d t}. 
\end{eqnarray}


Note that for phonons independently emitted by the TLS, $\bar{P}_{N}$ is simply 
proportional to the number $N$; $ \bar{P}_{N}=N\bar{P}$. 
However, in cooperative emission, the lifetime of the excited state of a 
single TLS is $N$ times shorter, hence the phonon emission power of a single TLS 
should be $N$ times larger when $\omega\tau_{\epsilon}>1$\cite{comment}.
This means that $ \bar{P}_{N}$ should be proportional to $N^{2}$ for $\omega\tau_{\epsilon}>1$.
Our numerical results
indeed show that the $\bar{P}_{N}$  has a quadratic dependence of $N$ (see Fig. 3.)

\begin{figure}
\includegraphics[width=8.0cm,height=7cm]{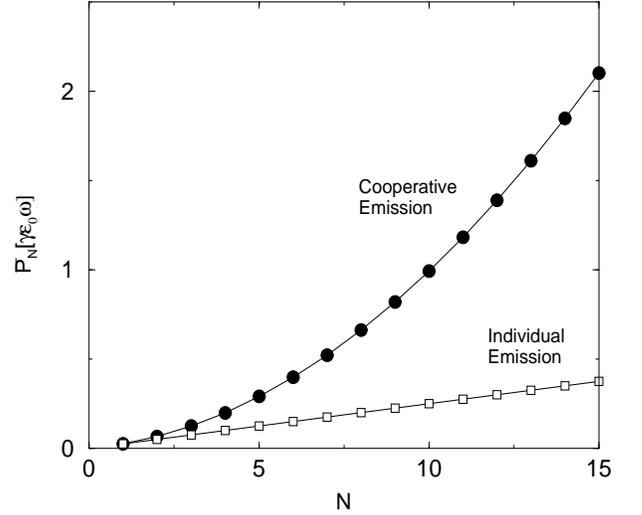}
 \caption{ The time-averaged phonon emission power $\bar{P}_{N}$ of $N$ symmetric TLSs
 ($\Delta_{s}=0$) at 
$T=0.1\gamma\epsilon_{0}/k_{B}$ and $\omega\tau_{\epsilon}=20$, which was numerically calculated using
Eq.(18).}
\end{figure}

When $N$ number of TLS are within $\lambda_{ph}$, one obtains
 $Q_{1}^{-1}$ by replacing $\bar{P}$ with its effective value $\bar{P}^{*}=\bar{P}_{N}/N$.
The enhanced phonon emission power due to superradiance $\bar{P}^{*}\sim N \bar{P}$
then provides a natural explanation of the anomalously-large internal friction at low temperatures. 
Including the phonon superradiance effect, Eq.(\ref{qq1}) for $\omega\tau_{\epsilon}>1$ can be rewritten as
\begin{eqnarray}
Q^{-1}(T=0)=Q_{1}^{-1}(T=0)\sim \frac{1}{\omega\tau_{\epsilon}^{*}}\frac{ n\gamma^{2}}{\rho v^{2}}, 
\end{eqnarray}
where $\tau_{\epsilon}^{*} \approx \tau_{\epsilon}/N$ is the renormalized relaxation time and $N$ is the number of cooperating TLSs. 
Thus, the enhancement of $Q^{-1}$ at low temperature by a factor of $10^{3}-10^{4}$ compared to 
our initial estimate in the independent TLS model
(Eq.(\ref{qq1})) indicates that a large number of $\sim 10^{3}-10^{4}$ TLS emit phonons cooperatively.
By taking $\lambda_{ph}\sim 10 \mu$m from $\lambda_{ph}=\frac{\hbar v_{s}}{\gamma \epsilon_{0}}$, 
the total number of TLSs in the volume $\lambda_{ph}^{3}$
is  $\sim 10^{5}$ according to Ref.\cite{phillips} or $\sim 10^{3}$ according to Ref.\cite{keyes}, which
is reasonably large enough number for the cooperative emission by $10^{3}$ TLSs.

In conclusion, we formulate a theory of internal friction of micro- and nano-mechanical resonators, 
which invokes intrinsic two-level defects in the nonlinear regime at high frequencies.
As temperature decreases, the mechanical friction crosses over from the linear regime 
to the nonlinear regime, resulting in the saturation behavior as $T$ goes to zero.
 Because of the phonon superradiance, the low-temperature saturation value of $Q^{-1}$ is 
strongly enhanced by a factor given by the number of two-level systems contained within the
emitted phonon wavelength.

\begin{acknowledgements}
This work was supported by Brain Korea 21 project. KHA is thankful to the Boston University Physics Department
 for the kind hospitality during his visit, where a part of this work was performed. PM thanks the
Dean's Office of Boston University for the support of this work.
\end{acknowledgements}


\begin{thebibliography}{99}
\bibitem{comment} Since the energy gap of our TLS has time variation, if $\omega\tau_{\epsilon} <1 $, 
reduction of the relaxation time gives rise to red shift of emitted phonon.
 So the phonon emission power of $N$ number of TLS is not simply proportial to $N^{2}$.
\end{thebibliography}
\end{document}